\documentclass[twocolumn,prl,superscriptaddress]{revtex4}
\pdfoutput=1
\usepackage{graphicx}
\usepackage[utf8]{inputenc}
\usepackage[T1]{fontenc}
\newcommand{\be}{\begin{equation}}
\newcommand{\ee}{\end{equation}}
\newcommand{\bea}{\begin{eqnarray}}
\newcommand{\eea}{\end{eqnarray}}

\newcommand{\la}{\langle}
\newcommand{\ra}{\rangle}

\renewcommand{\phi}{\varphi}
\renewcommand{\epsilon}{\varepsilon}
\renewcommand{\vec}[1]{{\bf #1}}

\begin{document}

\title{Nature of the spin liquid state of the Hubbard model on honeycomb lattice}
\author{B. K. Clark}
\affiliation{Princeton
  Center for Theoretical Science, Princeton University, Princeton, NJ
  08544} \affiliation{Department of Physics, Joseph Henry
  Laboratories, Princeton University, Princeton, NJ 08544}
\author{D. A. Abanin}
\affiliation{Princeton
  Center for Theoretical Science, Princeton University, Princeton, NJ
  08544} \affiliation{Department of Physics, Joseph Henry
  Laboratories, Princeton University, Princeton, NJ 08544}
\author{S. L. Sondhi}
\affiliation{Department of Physics, Joseph Henry
  Laboratories, Princeton University, Princeton, NJ 08544}

\date{\today}
\begin{abstract}
Recent numerical work (Nature 464, 847 (2010)) indicates the existence of a spin liquid phase (SL) that intervenes between the antiferromagnetic and semimetallic phases of the half filled Hubbard model on a honeycomb lattice. 
To better understand the nature of this exotic phase, we study the quantum $J_1-J_2$ spin model on the honeycomb lattice, which provides an effective description of the Mott insulating region of the Hubbard model. 
Employing the variational Monte Carlo approach, we analyze the phase diagram of the model, finding a phase transition between antiferromagnet and an unusual $Z_2$ SL state at $J_2/J_1\approx 0.08$, which we identify as the SL phase of the Hubbard model. At higher $J_2/J_1\gtrsim 0.3$ we find a transition to a dimerized state with spontaneously broken rotational symmetry.  
\end{abstract}

\maketitle


%









{\bf Introduction.}
The Hubbard model describes electrons hopping on a lattice and interacting via on-site Coulomb interactions, 
\be\label{eq:hubbard}
H=-t\sum_{\langle ij\rangle, s} a_{is}^\dagger a_{js}+U\sum_{i} n_{i\uparrow}n_{i\downarrow},
\ee
where $s=\uparrow,\downarrow$ denotes spin and $n_{is}=a_{is}^\dagger a_{is}$.  
Despite its conceptual simplicity, the Hubbard model exhibits a rich phase diagram and is believed to capture the physics of the high-temperature cuprate superconductors~\cite{Anderson87} (for a review, see Ref.~\cite{NagaosaWenLee}). 
At half-filling and strong repulsion, $U\gg t$, the Hubbard model is a Mott insulator, in which electrons are localized by strong Coulomb repulsion. The Mott insulator is characterized by a charge gap of the order $U$, and, in the 
limit $t\ll U$ the low-energy dynamics of this phase is associated with the spin degree of freedom. The effective spin-spin interactions, 
which originate from virtual hopping processes and intertwine spins of neighboring electrons, become increasingly frustrated as the ratio $U/t$ is lowered and the Mott transition is approached. 
In the vicinity of the Mott transition, the frustration enhances quantum fluctuations, which can prevent ordering of spins down to zero temperature, giving rise to spin liquid ground states. 
The interest in spin liquids stems from the fact that some of them exhibit new types of topological order~\cite{Wen89}, and the fact that their properties 
may be linked with the physics of the doped Hubbard model~\cite{NagaosaWenLee}.  

\begin{figure}
\includegraphics[width=3in]{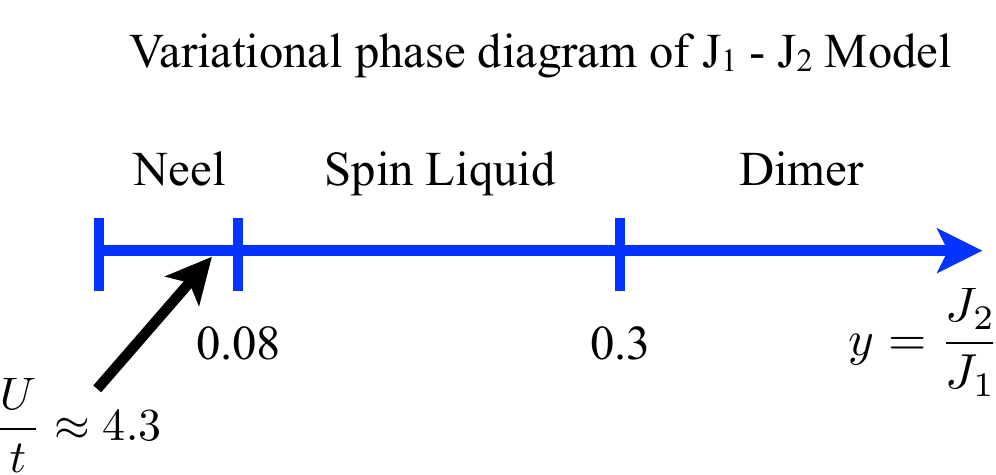}
\vspace{-1mm}
\caption[]{Phase diagram of the quantum $J_1-J_2$ model, obtained using the variational Monte Carlo method.  
At $J_2/J_1\approx 0.08$ we have found a phase transition between an antiferromagnetic state, realized in a pure Heisenberg model, and a gapped spin liquid state. The arrow indicates the location for this transition in the Hubbard model \cite{Assaad10}. The SL state is best described variationally by SPS. The spin model at $J_2/J_1 \lesssim 0.12$ describes the Mott insulating phase of the Hubbard model. At higher $J_2/J_1>0.3$ the SL gives way to a dimerized phase. The trial wave functions of  AFM states were of the Huse-Elser type~\cite{HuseElser}, and for SL and dimerized state of the RVB type~\cite{Gros89}.}
\label{fig1}
\end{figure}

\begin{figure}
\includegraphics[width=3.4in]{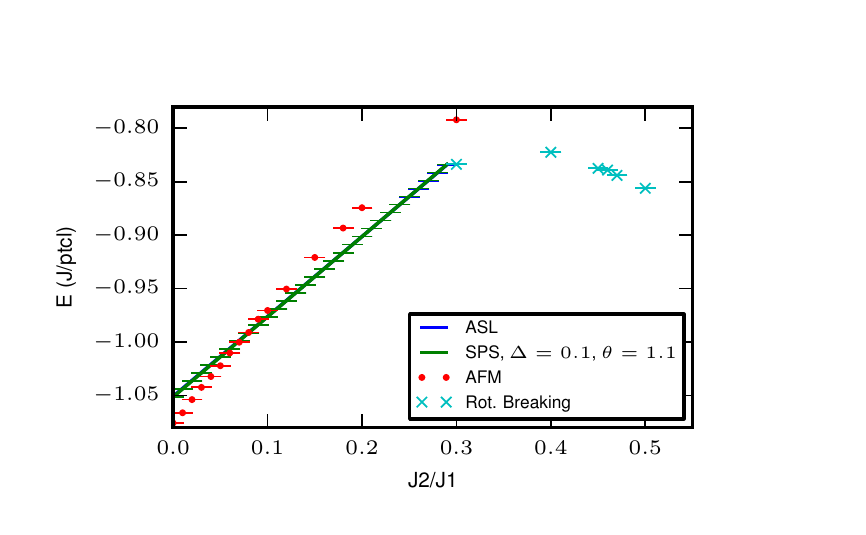}
\vspace{-1mm}
\caption[]{Energies of AFM, ASL, SPS, and dimerized phases compared for $10 \times 10$ system. 
AFM state is favorable at $J_2/J_1\lesssim 0.08$; 
ASL and SPS states have energy lower than AFM at $J_2/J_1\gtrsim 0.08$, but their energies are very close. 
Spontaneous breaking of the rotational symmetry occurs at $J_2/J_1\approx 0.3$, giving rise to the dimerized state.}
\label{fig2}
\end{figure}

\begin{figure}
\includegraphics[width=3.4in]{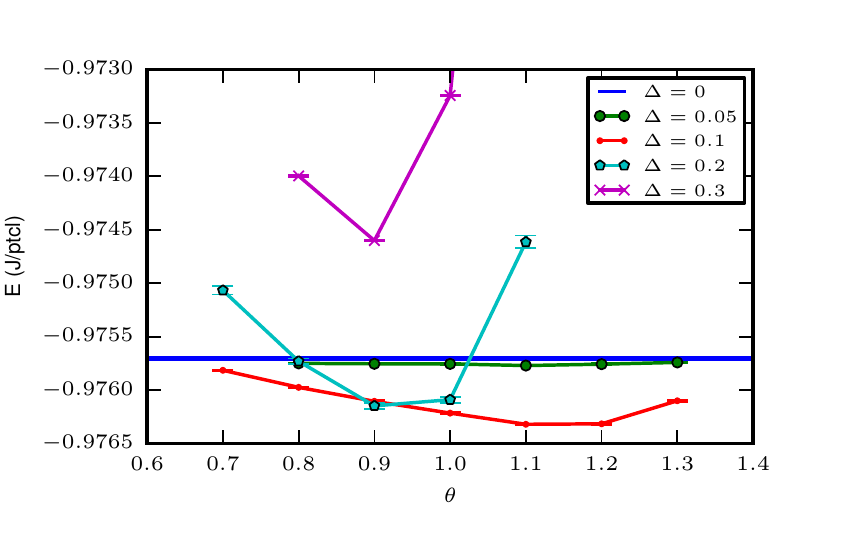}
\vspace{-1mm}
\caption[]{Energies of the SPS state for different values of pairing amplitude $\Delta$ as a function of pairing phase $\theta$ compared to the ASL energy (for system size $14\times 14$). SPS state is favored, and its energy is minimized for $\Delta\approx 0.1,\theta \approx 1.1$. 
}
\label{fig3}
\end{figure}

Recently, the Hubbard model on the honeycomb lattice at half-filling was studied using the determinantal quantum Monte Carlo (DQMC)  method~\cite{Assaad10}. 
The Hubbard model on the honeycomb lattice has a crucial advantage of being free of the sign problem, and therefore DQMC gives essentially exact results 
for correlators of the system. 
It was found that in the vicinity of the Mott transition, 
the system exhibits 
a disordered spin phase. This phase intervenes between the antiferromagnetic Neel state realized  at higher $U/t\approx 4.3$, and the semimetallic phase at $U/t <3.5$. 
The authors of Ref.~\cite{Assaad10} found that the disordered phase shows a small but finite spin gap, and preserves translational symmetry and time-reversal symmetry. This suggests that the disordered state 
on the honeycomb lattice is a non-chiral spin liquid.  

In this paper, we elucidate the nature of the spin liquid state (SL) on the honeycomb lattice. We study the effective $J_1-J_2$ spin model of the large-$U$ Hubbard model. Using Variational Monte Carlo (VMC) method, we find a phase diagram (FIg.1), which leads us to identify the exotic phase seen in the Hubbard model as the sublattice-pairing-state (SPS), a small-gap $Z_2$ SL, first considered in Ref. [7]. Our results should be contrasted with the mean-field analysis of Ref.[7] which favors a gapless SL, rather than SPS, in the relevant parameter range.
We attribute the difference to the fact that the mean-field approach [7] neglects essential gauge fluctuations, which are accounted for by VMC.

{\bf Spin Hamiltonian}
We start from an effective spin Hamiltonian, the $J_1-J_2$ spin model,
\be\label{eq:model}
H=J_1\sum_{\la ij \ra} {\bf S}_i \cdot {\bf S}_j+J_2\sum_{\la\la ij \ra\ra}{\bf S}_i \cdot {\bf S}_j,
\ee
where $\la  ij\ra$, $\la\la ij \ra\ra$ denote nearest-neighbor and next-nearest-neighbor sites.  To establish the connection with the Hubbard model, we calculate the parameters of the spin model from the perturbation theory in $(t/U)^2$~\cite{MacDonald90}, finding, to second order in $(t/U)^2$,
\be\label{eq:parameters}
J_1=4\frac{t^2}{U}-16\frac{t^4}{U^3}, \,\, J_2=4\frac{t^4}{U^3}.
\ee
The SL phase in the Hubbard model then ranges from $J_2/J_1 \approx 0.07$  at the antiferromagnetic (AFM) transition
to $J_2/J_1 \approx 0.12$ at the semi-metal transition.
Both exchange couplings are antiferromagnetic, thus the effective spin model is frustrated. 
In this study we ignore higher order terms in $(t/U)^2$ as well as third nearest neighbor and ring exchange terms\cite{RingExchange}. At the 
relevant $U/t\approx 4.3$ higher order terms are likely too small to affect the results and third nearest neighbor terms are non-frustrating
and will just renormalize the effective $J_1$.

Quantum fluctuations are particularly important in this model as they are enhanced by the low cooridination number of the 
honeycomb lattice and competition between the nearest and 
next-nearest-neighbor exchange interactions. We contrast the phase diagram of the quantum model (Fig.~1) with that of the classical $J_1$ - $J_2$ model. 
The latter model exhibits two phases: Neel antiferromagnet, with opposite spin polarization on the two 
sublattices of the honeycomb lattice, and (an incommensurate) spiral ordering~\cite{honeycomb-classical}. 
The Neel state, which is favored by the nearest-neighbor coupling, is realized at small $J_2/J_1<1/6$, where the nearest-neighbor coupling dominates. 
When $J_2/J_1>1/6$, spiral order with an incommensurate wave vector sets in. The origin of the spiral order is especially clear at very large $y\gg 1$, when nearest-neighbor coupling is negligible, and the two triangular sublattices can be thought of as independent; the ground state on each sublattice is a spiral state in which spins on three sub-sublattices are rotated by $2\pi/3$ with respect to each other. 

We find the quantum fluctuations drastically alter this phase diagram. The spiral phase is destroyed at intermediate $J_2/J_1$, giving way to a 
disordered spin-liquid phase at $J_2/J_1<0.3$, and a dimerized phase at $J_2/J_1 \gtrsim 0.3$.  
The Neel state survives at small $J_2/J_1 \lesssim 0.08$, albeit with reduced magnetization.

{\bf Ans\"atze}. 
The main goal of this work is to understand the nature of the spin-liquid phase.  
Toward that end we focus on two primary types of wave-functions:
(generalized) Huse-Elser \cite{HuseElser,CPS} states and resonating valence bond (RVB) states \cite{Gros89}.  

The former of these is chosen as a good ansatz for the 
AFM state.
In the Huse-Elser wave function, the phase of the wave function is fixed by the Marshall
sign rule and for the real part we optimize a separate variational two-body parameter $C(\vec{r})$  for each unique vector $\vec{r}$.\


The RVB state is represented as 
\be\label{eq:RVB}
|\psi_{RVB}\ra=\sum_{\{ \cal D\}} {\cal A_{D}} \prod _{i,j} |\uparrow_i \downarrow_j -\downarrow_i \uparrow_j \ra
\ee  
where $\{ \cal D\}$ is a (generically non-nearest neighbor) dimer covering of the lattice. 
Different choices for ${\cal A_{D}}$ correspond to qualitatively different types of wave functions.  RVB states are
good ans\"atze for (gapped and gapless) spin-liquid states as well as dimer states.

One approach for selecting these amplitudes is to write down a large (but not complete) set of parameters specifying 
the RVB amplitudes 
and then optimize over them.  
We parameterize ${\cal A_{D}}$ so
as to be able to represent all states of the form 
$P \prod_k (u_k+v_k c_{k,\uparrow}^{\dagger} c_{-k,\downarrow}^{\dagger})\left|0\right>$ for any real $u_k,v_k$
where $P$ projects out double occupancy. We call these generic RVB states.  
Optimization for these (and the Huse-Elser) states is done via stochastic
optimization \cite{StochasticOpt}.  
This parameterization has two drawbacks.  First,
because optimization of a large
set of parameters runs the risk of being stuck in local minima, 
we are not guaranteed to find the best state.  Secondly, 
some spin liquid states, such as SPS (see below), are not 
encompassed by this parameterization. 
Because of these concerns, we also generate RVB amplitudes 
in a more physically motivated way 
allowing for fewer parameters that can be 
directly swept over. 

This alternative approach uses the Schwinger fermion representation of the 
spin model combined with Gutzwiller projection~\cite{Gros89}. 
In this approach, the spin operator on the $i$th site is related to fermionic creation-annihilation operators $f, f^\dagger$ as follows, $ S^\alpha_i=\sum_{s,s'}f^\dagger_{i s}\sigma^{\alpha}_{ss'} f_{is'}$, and a constraint of one fermion per site is imposed, $\sum_{s} f^\dagger_{is}f_{is}=1$. Wave functions of the spin model are obtained by Gutzwiller projection of the fermionic many-body wave functions, which projects out sites with double or no occupancies. 

The fermionic wave functions are then generated as ground states of a quadratic Hamiltonian on the honeycomb lattice 
\be\label{eq:hamiltonian1}
H_F=-t\sum _{\la ij \ra, s} f_{is}^\dagger f_{js} +\sum _{ij}  \Delta_{ij}( f^\dagger _{i \uparrow}f^\dagger_{j\downarrow}-f^\dagger _{i \downarrow}f^\dagger_{j\uparrow} )+h.c. 
\ee
which includes nearest-neighbor hopping, and superconducting pairing. In principle, a small second neighbor hopping 
should also be included as suggested by the RVB mean-field theory of this state but we ignore it in this first pass at the problem.  The parameters $\{\Delta \}$ are chosen in such a way that the ground state energy of 
the projected wave-function is minimized. 


An important advantage of the Schwinger fermion representation is that there exist simple choices of $\{ \Delta\}$, with just hopping matrix element between nearby neighbors, which describe different types of candidate spin liquid states that have been proposed. Lee and Lee~\cite{Lee05}, and later Hermele~\cite{Hermele08} conjectured the existence of an algebraic spin liquid (ASL) on the honeycomb lattice, which is characterized by gapless spin excitations with Dirac-like spectrum, similar to that in graphene.  This corresponds to the nearest-neighbor tight-binding model, with $\Delta_{ij}=0$ for all $i,j$.   Very recently, Lu and Ran~\cite{Ran10}  suggested two possible gapped spin liquid states, which differ by their projective symmetry group~\cite{Wen_symmetry}; we will call these states $s$-wave spin liquid ($s$SL), and sublattice-pairing state (SPS). $s$SL is obtained by considering real $\Delta_{ij}$'s, which are rotationally and translationally invariant, for sites $i,j$ which are $n$th nearest neighbors or closer. SPS is characterized by complex pairing amplitudes, with opposite phases on the two sublattices~\cite{Ran10},
\be\label{eq:SPS}
\Delta_{ij}=\Delta e^{i\theta},  i,j\in A, \quad \Delta_{ij}=\Delta e^{-i\theta}, \quad i,j\in B
\ee
where $i,j$ are next-nearest neighbors.
Within their RVB mean-field theory, Lu and Ran find that ASL is favored over $s$SL and SPS  when $J_2/J_1<0.3$ which encompasses the entire regime relevant to the spin-liquid in the Hubbard Model.
In all the spin-liquid ans\"atze, the  symmetries (translational, time-reversal, rotational symmetry) of the honeycomb lattice are respected. 


\begin{figure}
\includegraphics[width=3.4in]{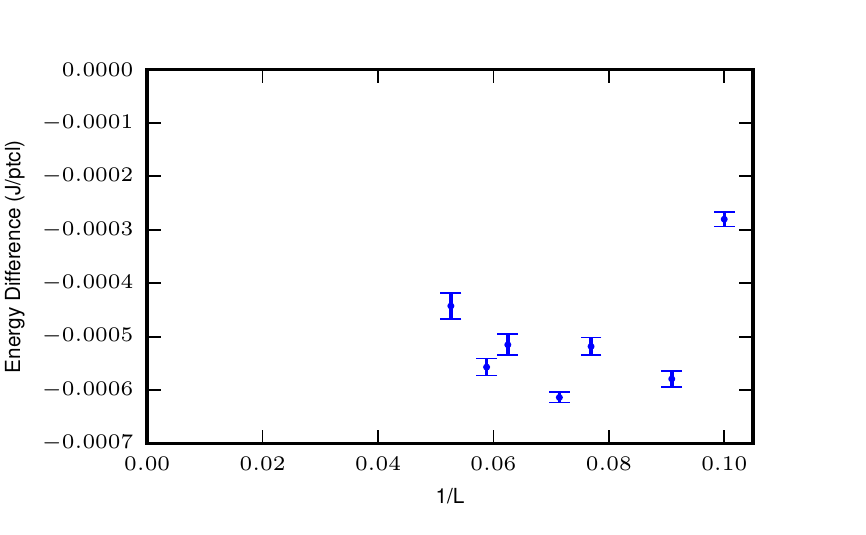}
\vspace{-1mm}
\caption[]{Energy difference between SPS state with optimized (for 14 x 14) pairing parameters ($\Delta\approx0.1, \theta\approx1.1$) and ASL state as a function of system size. The energy difference extrapolates to a non-zero value in the thermodynamic limit $N\to \infty$. Non-monotonicity of the points is a 
result of incommensurability effects with the lattice. }
\label{fig4}
\end{figure}

\begin{figure}
\includegraphics[width=3.4in]{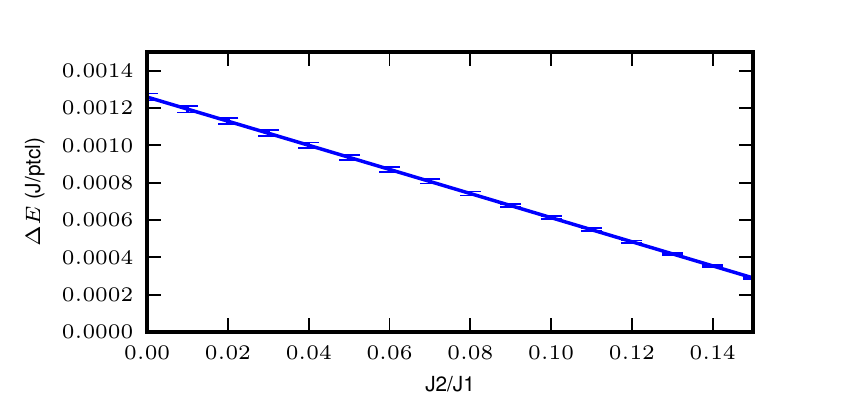}
\vspace{-1mm}
\caption[]{Energy difference for system size $14 \times 14$ between SPS ($\delta=0.1,\theta=1.1$) and ASL states, as a function of the frustration parameter $J_2/J_1$. $\Delta E>0$ indicate the SPS is variationally preferred. The energies of the two phases cross
at $J_2/J_1 \gtrsim 0.2$ and further study is warranted to determine if SPS with different parameters still dominates ASL
in this regime.}
\label{fig5}
\end{figure}

{\bf Results.}
Using the variational approach, we have mapped out the energies of various phases, illustrated in Fig.~\ref{fig2}. 
We find a phase transition between AFM and SL at $J_2/J_1\approx 0.08$. Both phases are also found in the Hubbard model~\cite{Assaad10}, 
and the transition point is remarkably close to that in the Hubbard model.  At higher frustration 
parameters ($J_2/J_1 >0.3$), we find the rotational symmetry of the RVB states is broken
giving a dimerized state a lower energy then that of the SL phase.
This is seen by optimizing RVB amplitudes up to third nearest neighbors 
(with the other amplitudes fixed as in the ASL state)
and is consistent with findings from exact diagonalization studies
on small clusters (\cite{Fouet01,Mosadeq00}) which have suggested dimerized states.
Having identified the location for the spin-liquid phase, we turn our 
attention to identifying its nature.  
 
In establishing the form of spin liquid state, we focus on the ASL and the 
SPS state, which is variationally the lowest gapped state we find.   We do not consider $s$SL as our optimization over generic RVB states (which includes $s$SL) does not find a lower state then SPS.  We notice that the energy difference between ASL and SPS is very small, and a more careful study is needed to distinguish between them. 

To establish whether SPS is more favorable than the ASL, we have optimized the SPS energy with respect to pairing amplitude and phase. We first consider $J_2/J_1=0.1$, and a $14\times 14$ system. By mapping out the energy as a function of $\Delta,\theta$ (see Fig.~\ref{fig3}), we have established the optimal values $\Delta/t\approx 0.1$, $\theta\approx 1.1$.

The energy of the SPS state with those parameters is lower than that of ASL, suggesting that ASL is unstable with respect to pairing that opens a gap. However, the energy gain due to the gap opening is so small, that one may doubt whether it survives in the thermodynamic limit. To answer this question, we
studied finite-size scaling of the energy difference $E_{SPS}-E_{ASL}$ at the parameter values $\Delta/t=0.1,\theta=1.1$ 
 considering $N=\{10,11,13,14,16,17,19\}$.
The result, illustrated in Fig.~\ref{fig4}, clearly shows that the energy difference extrapolates to a non-zero value in the thermodynamic  limit $1/N\to 0$, indicating that SPS is the ground state.

\begin{figure}
\includegraphics[width=2in]{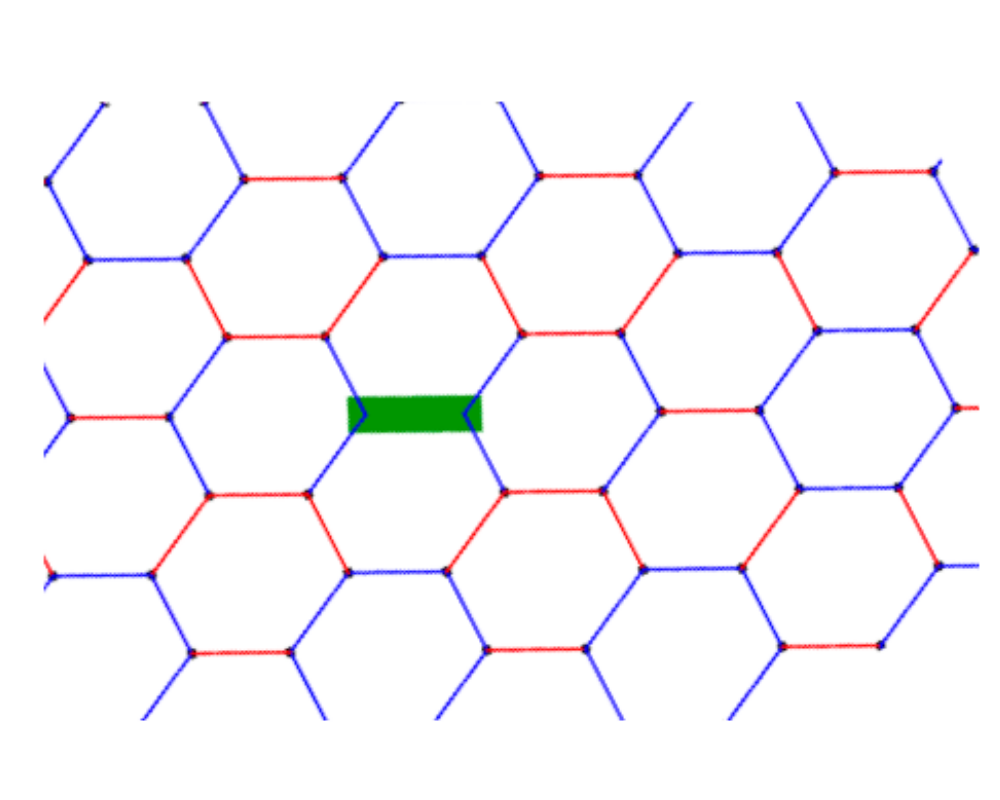}
\vspace{-1mm}
\caption[]{Dimer-dimer correlation function, defined as in
\cite{Assaad10} for a SPS 
state. Green is the reference slice.  Red indicates a positive correlation with the reference slice and blue a negative correlation.  
The SPS has the same positive and negative correlations as the dimer-dimer correlations found in \cite{Assaad10}. 
}
\label{fig6}
\end{figure}

We have repeated the comparison between energies for ASL and SPS in the whole range of frustration parameter $0.05<J_2/J_1<0.25$; the result for the $14\times 14$ system is illustrated in Fig.~\ref{fig5}. We found that SPS state is favored in the range $0.05<J_2/J_1<0.2$, and at larger values of the frustration parameter the energies of the two phases are swapped. Additionally at these higher frustration parameters, we find a generic RVB state that does not break sublattice symmetry and has a lower energy then the ASL state.  Because we have not studied the finite-size effects or optimized carefully the SPS parameters at these higher frustration parameters, this could either point to a series of phase transition between the different states, or to the SPS gap becoming too small to be resolved without more careful optimization and finite size extrapolation. Further work is needed to distinguish between these different scenarios.

Having identified SPS as the variationally lowest energy state, we examine some of its properties.
We start by naively estimating the gap.
We accomplish this by  assuming that the gap corresponds to the BCS pairing gap
of the tight-binding model obtained by a mean-field decoupling of the 
Schwinger fermion model.  For the $14\times 14$ system we find
\be
\Delta E =\frac{\pi}{2}\frac{\Delta}{t} J_1\cos\theta \approx \frac{t}{20},
\ee
which compares favorably with Hubbard model result \cite{Assaad10}, $\Delta E \approx t/13$ for the same system size (in the thermodynamic limit $\Delta E \to t/40$). 

We also look at the dimer-dimer correlation function for an AFM and SL state (see Fig.~\ref{fig5}).  We find that the dimer-dimer correlations of the SL state are positively (respectively negatively) correlated  on exactly the same dimers as the Hubbard model at $U/t \approx 4.0$ \cite{Assaad10}. It should
be noted that although the ASL state has a similar looking dimer pattern, due to the small gap of the SPS, the AFM state looks qualitatively different. 

 
 {\bf Discussion.} In conclusion, we have studied $J_1-J_2$ model on the honeycomb lattice, finding evidence for SL phase in the range $0.08<J_2/J_1<0.3$. 
We have accumulated strong evidence that the SPS state describes 
the spin-liquid phase seen in the Hubbard Model.  
Beyond having the transition happen in almost exactly the correct
place, we find it to be the variational 
lowest energy state beating out the gapless ASL state.
Additionally, in naively estimating the gap, we find that it corresponds 
well to the Hubbard gap.  Moreover, the dimer-dimer correlations 
closely match those of the Hubbard model. We should note that for $SU(2)$, as opposed to large $N$ $SU(N)$ spins, it is not clear whether the SPS state represents a phase of matter that is distinct from the simplest short ranged RVB phase obtained, e.g., in the quantum dimer model~\cite{MoessnerSondhi}.  

{\bf Acknowledgements.} We thank Duncan Haldane, David Huse  and M.P.A. Fisher for many invaluable discussions. DA thanks Aspen Center for Physics, where part of this work was completed for hospitality during program ``Topological phases''. 



\begin{references}
\bibitem{RingExchange}
Because the smallest ring exchange term involves six sites, we 
suspect that  large energy differences are almost
certainly robust against it. Small energy differences (as we find between different spin liquids) may be sensitive to its presence.
\end{references}
\end{document}